\documentclass[journal=jacsat,manuscript=article]{achemso}

\usepackage[version=4]{mhchem}
\usepackage{nicefrac}
\usepackage{siunitx}
\usepackage{graphicx}

\author{Emanuele Grifoni}
\affiliation[ETH]
{Department of Chemistry and Applied Biosciences, ETH Zurich, c/o USI Campus, Via Giuseppe Buffi 13, CH-6900 Lugano, Ticino, Switzerland}
\alsoaffiliation[USI]
{Institute of Computational Science, Università della Svizzera italiana (USI), Via Giuseppe Buffi 13, CH-6900, Lugano, Ticino, Switzerland}
\author{GiovanniMaria Piccini}
\affiliation[ETH]
{Department of Chemistry and Applied Biosciences, ETH Zurich, c/o USI Campus, Via Giuseppe Buffi 13, CH-6900 Lugano, Ticino, Switzerland}
\alsoaffiliation[USI]
{Institute of Computational Science, Università della Svizzera italiana (USI), Via Giuseppe Buffi 13, CH-6900, Lugano, Ticino, Switzerland}
\author{Michele Parrinello}
\email{parrinello@phys.chem.ethz.ch}
\affiliation[ETH]
{Department of Chemistry and Applied Biosciences, ETH Zurich, c/o USI Campus, Via Giuseppe Buffi 13, CH-6900 Lugano, Ticino, Switzerland}
\alsoaffiliation[USI]
{Institute of Computational Science, Università della Svizzera italiana (USI), Via Giuseppe Buffi 13, CH-6900, Lugano, Ticino, Switzerland}
\alsoaffiliation[IIT]
{Italian Institute of Technology, Via Morego 30, 16163 Genova, Italy}

\title[]
  {A microscopic description of acid-base equilibrium} 


\begin{document}

%
%
%
%
%

\begin{abstract}
Acid-base reactions are ubiquitous in nature. Understanding their mechanisms is crucial in many fields, from biochemistry to industrial catalysis. Unfortunately, experiments only give limited information without much insight into the molecular behaviour. Atomistic simulations could complement experiments and shed precious light on microscopic mechanisms. The large free energy barriers connected to proton dissociation however make the use of enhanced sampling methods mandatory. Here we perform an ab initio molecular dynamics (MD) simulation and enhance sampling with the help of methadynamics. This has been made possible by the introduction of novel descriptors or collective variables (CVs) that are based on a conceptually new outlook on acid-base equilibria. We test successfully our approach on three different aqueous solutions of acetic acid, ammonia, and bicarbonate. These are representative of acid, basic, and amphoteric behaviour. 
\end{abstract}

\section{Introduction}
Acid-base reactions play a key role in many branches of chemistry.
Inorganic complexation reactions, protein folding, enzimatic processes, polymerization, catalytic reactions and many other transformations in different areas are sensitive to changes in pH. Understanding the pH role in these reactions implies having control over their reactivity and kinetics.

The crucial importance of pH has stimulated the collection of a large amount of data on acid-base equilibria. These are typically measured in gas and condensed phases using spectroscopic and potentiometric techniques. However, there are practical limitations to the accuracy of these methods especially in condensed phases\cite{Ho2011}. Furthermore it is very difficult to extract from experimental data a microscopic picture of the processes involved. It is thus not surprising that acid-base equilibrium has been the subject of intense theoretical activity\cite{Elstner2001,Saracino2003AbsoluteModel,Schuurmann1998PredictionPCM-UAHF,Ho2011,Ho2009,Silva2000,Rebollar-Zepeda2016,Davies2002,Park2006,Tummanapelli2014,Ortiz2018,Lee2006}.

The acidity of a chemical species in water can be expressed in terms of $pK_a$, the negative logarithm of the acid dissociation constant.
There are two ways of calculating these values, one static and the other dynamic.

The most standard approach is the static one in which solution-phase free energies, and consequently $pK_a$s, are obtained closing a Born-Haber cycle composed by gas phase and solvation free energies\cite{Saracino2003AbsoluteModel,Schuurmann1998PredictionPCM-UAHF,Ho2011,Ho2009,Silva2000,Rebollar-Zepeda2016}. While extremely successful in many cases, the static approach has some limitations. A solvation model needs to be chosen and continuum solvent models have a limited accuracy. This is particularly true in systems like zeolites or proteins characterized by irregular cavities in which an implicit description of the solvent is challenging. Obviously from such an approach dynamic information cannot be gained. Furthermore, there can be competitive reactions that cannot be taken into account unless explicitly included in the model.

In principle these limitations could be lifted in a more dynamical approach based on MD simulations in which the solvent molecules are treated explicitly. If one had unlimited computer time such simulation would explore all possible pathways and assign the relative statistical weight to the different states. Unfortunately the presence of kinetic bottlenecks frustrates this possibility trapping the system in metastable states, since different protonation states are separated by large barriers. Furthermore in acid-base reactions chemical bonds are broken and formed. This requires the use of ab initio MD in which the interatomic forces are computed on the fly from electronic structure theories. This makes the calculation more expensive and reduces further the time scale that can be explored. 

To overcome this difficulty, the use of enhanced sampling methods \cite{Bernardi2015} that accelerate configurational space exploration becomes mandatory. A very popular class of enhanced sampling methods is based on the identification of the degrees of freedom that are involved in the slow reaction of interest. These degrees of freedom are usually referred to as collective variables (CVs) and are expressed as explicit functions of the atomic coordinates $\mathbf{R}$. Sampling is then enhanced by adding a bias that is a function of the chosen CVs\cite{Laio2002,Valsson2014,Torrie1977}. Furthermore, designing a proper set of good CVs has also a deeper meaning. Successful CVs capture in a condensed way the physics of the problem, identify its slow degrees of freedom and lead a useful modellistic description of the process.

In standard chemical reactions, this is relatively simple since well defined structures can be assigned to reactants and products\cite{Piccini2017,Mendels2018,Piccini2017a}. This is not the case for acid-base reactions in which a proton is added to or subtracted from the solute. Once this process has taken place, water ions (\ce{H+} or/and \ce{OH-}) are solvated and their structure becomes elusive. In fact water ions can rapidly diffuse in the medium via a Grotthuss mechanism\cite{Agmon1995}. They became highly fluxional and the identity of the atoms taking part in their structure changes continuously. The nature of these species is thus difficult to capture in an explicit analytic funcion of $\mathbf{R}$. However, given the relevance of acid-base reactions, many attempts have been made at defining these entities\cite{Davies2002,Park2006,Tummanapelli2014,Ortiz2018,Lee2006}. Unfortunately these CVs have an ad-hoc nature and, while successful in this or that case, cannot be generally applied.

In order to build general and useful CVs we make two conceptual steps. One is to look at the acid-base process as a reaction involving only a few moieties. Namely the whole solvent and the reacting residues in the solvated molecule. For example when there is only one type of dissociating residue we think of the acid-base equilibrium as a reaction of the type
\begin{equation}
   \ce{A + H_{2N}O_N <=> B^{q_0} + H_{2N+q_1}O_N^{q_1}},
\end{equation}
where $N$ is the number of water molecules, $A$ and $B$ are a generic acid-base molecule in solution and its conjugate species respectively, $q_{0}$ and $q_{1}$ are integers that can assume values $+1$ and $-1$ according to the acid-base behaviour of the species and $q_1 + q_0 = 0$.

This implies that we do not look at the solvent as a set of molecules that compete to react with the acid-base species. Rather we consider the solvent in its entirety as one of the two adducts. Taking this point of view is especially relevant in polar solvents like water that are characterized by highly structured networks. In this case the presence of an excess or a deficiency  of protons changes locally the network structure and this distortion propagates along the entire network. 

Since the very early days of Eigen and Zundel\cite{Zustand1907,Uberschub-protonen1968}, researchers have struggled with how many molecules should be included in the definition of the perturbation\cite{Marx1999,Hulthe1997,Iyengar2005}. Given the absence of physical parameters capable of giving a clear and unequivocal answer to this question, the idea of considering the solvent as a whole circumvents this problem. Thus the solvent is not just a medium with a passive role, but it is looked at as an ensemble of molecules that contribute collectively to the formation of the conjugate acid-base pair. This point of view is much closer to the original one proposed by Br{\o}nsted and Lowry in which the reaction can be seen as a simple exchange of an hydrogen cation between an acid-base pair.

\begin{figure}
\centering
\includegraphics[width=.7\linewidth]{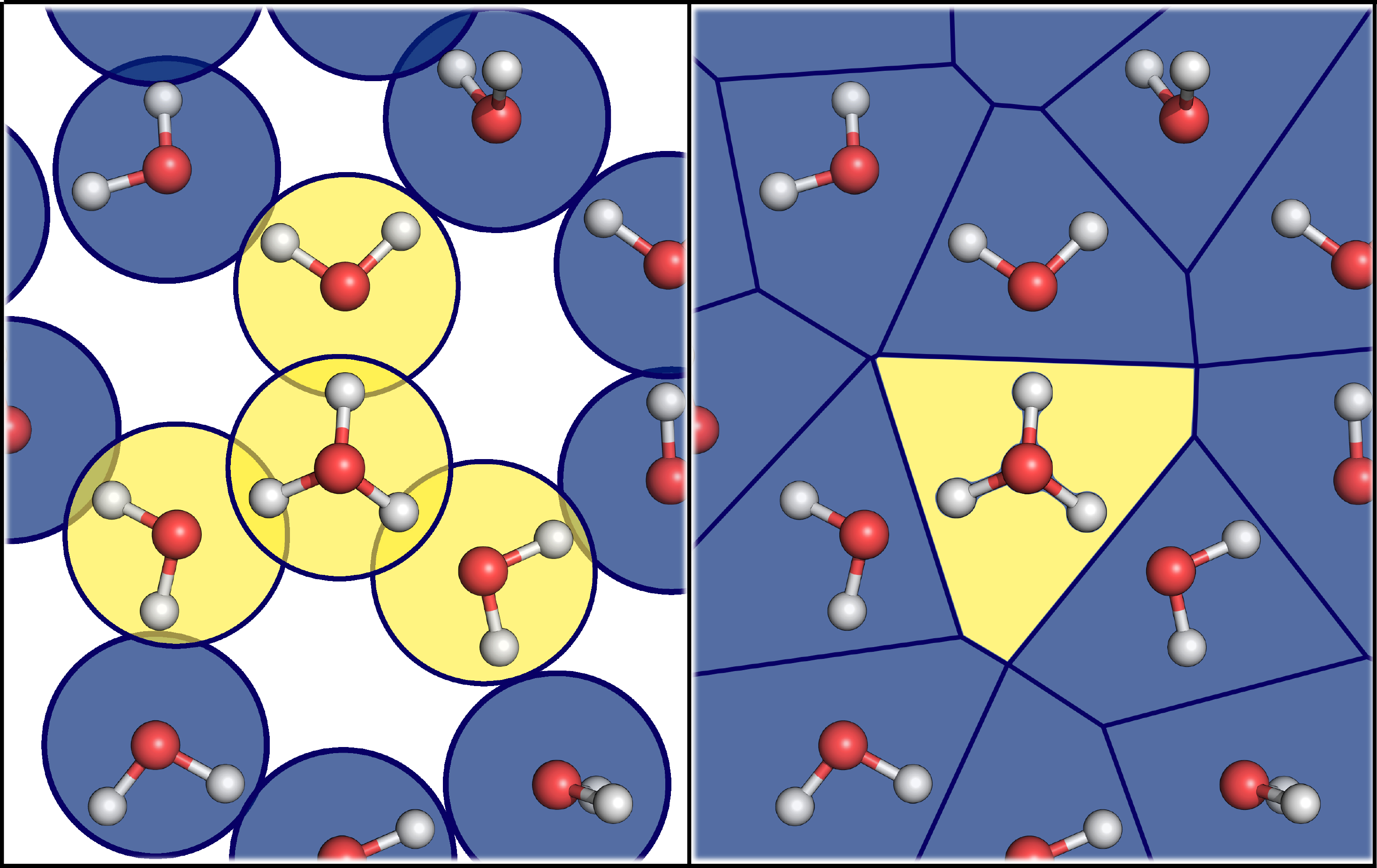}
  \caption{Two examples of partitioning the space. On the left we show a convectional approach in which the distance from oxygen atom is used to define its surrounding. Clearly artificial superpositions can be seen. On the right the Voronoi tessellation does not suffer from these shortcomings.} 
  \label{fig:voronoi+cutoff}
\end{figure}

For the reaction to take place the center of the perturbation has to move away from the solute. Thus the second important step is to monitor the center of the perturbation. Due to Grotthuss-like mechanisms the perturbation moves along the network. This can lead to different definitions of the defect center. However, if we tessellate the whole space using Voronoi polyhedra centered on water oxygen atoms we can assign unequivocally every hydrogen atom to one and only one of these polyhedra. The site whose Voronoi polyhedron contains an anomalous number of protons is taken as the center of the perturbation (see Fig.~\ref{fig:voronoi+cutoff}). 

This new point of view gives the method a very general nature making it applicable to every acid-base system, without the need of fixing beforehand the reacting pairs. Thus it is possible to explore all the relevant protonation states even in systems composed by more than one acid-base pair. 

This general approach allows defining CVs without having to impose specific structures or select the identity of the atoms involved. We test our method by performing metadynamics simulations in a weak acid case (acetic acid), a weak base (ammonia) and in an amphoteric species (bicarbonate) chosen as benchmark because of their comparable strength, but different acid-base behaviour. 

\section*{Methods}
\begin{figure}
\centering
\includegraphics[width=0.5\linewidth]{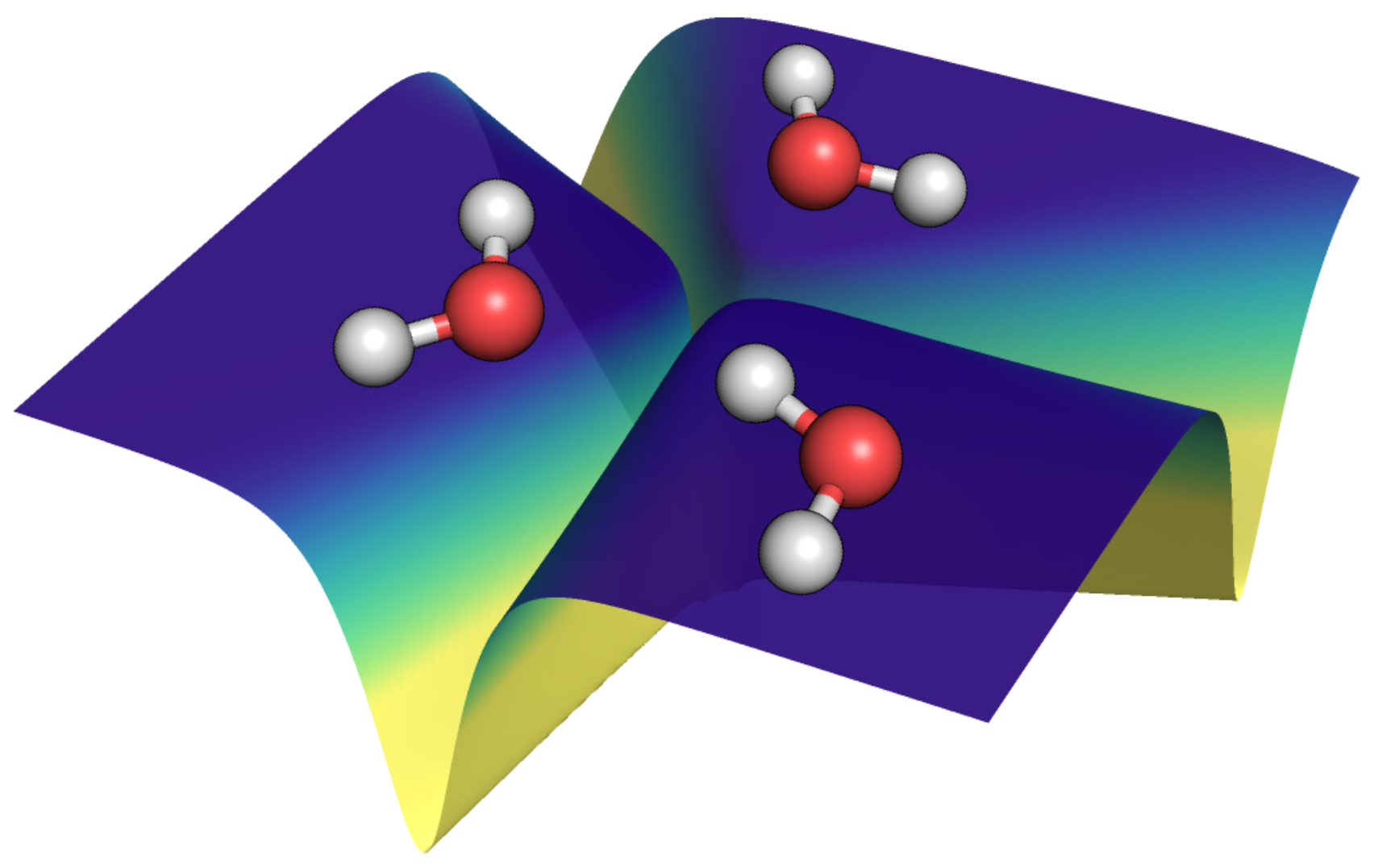}
  \caption{Smooth tessellation of a 2D space with cells centered on the 3 water molecule oxygen atoms. The flat blue regions represent the portion of space in which the function assumes a value of 1 and the yellow ones represent the borders among cells. This surface has been obtained with a value of $\lambda=4$.}
  \label{fig:surf}
\end{figure}

As discussed above we introduce two CVs, one related to the protonation state and the other that locates the charge defects and measures their relative distance.
Both of these CVs need a robust definition for assigning the hydrogen atoms to the respective acid-base site.
In order to achieve this result we partition the whole space into Voronoi polyhedra centered on the acid-base sites $i$ located at $R_i$. The sites include all the atoms able to breaking and forming bonds with an acid proton. The standard Voronoi space partition is described by a set of index functions $w_i(r)$ centered on the different $R_i$s such that $w_i(r)=1$ if the $i$-th atom is the closest to $r$, and equal to 0 otherwise. For their use in enhanced sampling methods CVs need to be differentiable. To this effect we introduce a smooth version of the index functions, $ w_i^s(r)$. These are defined using softmax functions:

\begin{equation}
  w_i^s(r)=\frac{e^{-\lambda |R_i-r|}}{\displaystyle\sum_m e^{-\lambda |R_m - r|}}, 
\end{equation}
where $i$ and $m$ run all over the acid-base sites and $\lambda$ controls the steepness with which the curves decays to 0, that is the selectivity of the function. With an appropriate choice of $\lambda$ this definition achieves the desired result as shown in Fig.~\ref{fig:surf}. In such a way, an hydrogen atom in a position $R_j$ is assigned to the polyhedron centered on the site $i$ with the weight $w_i(R_j)$. Then, the total number of hydrogen atoms assigned to the $i$-th acid-base site is:

\begin{equation}
  W_i=\sum_{j \in H} w_i^s(R_j),
\end{equation}
where the summation on $j$ runs all over the hydrogen atoms. 

One can associate to each acid-base site a reference value $W_i^0$ that counts the number of bonded hydrogen atoms in the neutral state. The difference between the instantaneous value of hydrogen atoms and the reference one is

\begin{equation}
    \delta_i=W_i-W_i^{0}.
\end{equation}
When different from zero $\delta_i$ will signal whether the $i$-th site has gained or lost a proton. 
In the case of water oxygen atoms, a hydronium ion has a $\delta_i=+1$ while a hydroxyde ion has $\delta_i=-1$.

We then group the acid-base sites in species. For instance in the case of the simplest amino acid glycine in aqueous solition the number of species $N^s$ will be equal to 3. All water oxygen atoms belong to one species, then one counts in another species the two carboxylic oxygen atoms and finally one considers as the third species the nitrogen atom of the amino group.

In the spirit of this work we count the total excess or defect of proton associated to each species,

\begin{equation}
    q_k = \sum_{i \in k} \delta_i.
    \label{eq:q}
\end{equation}

This implies that we are not interested in the specific identity of the reacted site, but whether or not the $k$-th species in its entirety has gained ($q_k=+1$), lost ($q_k=-1$) or has not changed its number of protons. 
If we consider a solute with only one reactive moiety then each possible state of the system can be described by one of the three two dimensional vectors (0,0), (-1,1) or (1,-1). 

In the general case each protonation state can be described by a vector $\Vec{q}=(q_0,q_1,\dots q_{N^s-1})$ with dimension equal to the number of inequivalent reactive sites, $N^s$. A more exhaustive explanation is provided in the S.I.

For use in enhanced sampling these vectors need to be expressed as a scalar function $f=f(\Vec{q})$ such that, for each physically relevant $\Vec{q}$, $f$ attains values able to distinguish the different overall protonation states. There are infinite many ways of constructing a scalar from a vector. Possibly the simplest choice is to write $f(\Vec{q})=\Vec{X} \cdot \Vec{q}$ and, in order to distinguish between different protonation states, to choose $\Vec{X}=(2^0,2^1,2^2,\dots 2^{N^s-1})$.

This leads to the following definition for the CV, that is used to describe the protonation state of the system: 
\begin{equation}
    s_p = \sum_{k=0}^{N^s -1} 2^k \cdot q_k,
\end{equation}
where $k$ are the indexes used to label the respective reactive site groups. In the appendix an example is worked out in detail. Of course the CV is made continuous by the use of $W_i$ in the calculation of the $\delta_i$ needed to evaluate $q_k$ in Eq.~\ref{eq:q}.

The second CV is a summation of distances between every acid-base sites multiplied for their partial charge $\delta_i$. 
\begin{equation}
    s_d = \sum_{i, m > i} -r_{im} \cdot \delta_i \cdot \delta_m,
    \label{eq:s_d}
\end{equation}
where the indexes $i$ and $m$ run all over the acid-base sites belonging to different $k$ groups, and $r_{im}$ is the distance between the two atoms. In this way, just the acid-base pair that has exchanged a proton gives a contribution different from zero. Eq.~\ref{eq:s_d} is valid only when one single conjugate acid-base pair is present. However, due to the action of bias it may occur occasionally that several acid-base pairs may be formed. In order to avoid sampling these very unlikely events we apply a restraint on the number of pairs. Further details are provided in the S.I.

\section*{Results}

We have applied our method to three aqueous solution of acetic acid, ammonia and bicarbonate as representations of a weak acid, a weak base and an amphoteric compound respectively. The setup of all three simulation is identical except for the identity of the solvated molecules. This ensures that the outcome reflects the different chemistry of these three systems and that there is no bias due to the initial condition.

Each simulation of the systems has been performed with Born-Oppenheimer MD simulations combined with well-tempered metadynamics \cite{Laio2002,Barducci2008} using CP2K package \cite{Vandevondele2005} patched with PLUMED 2 \cite{Brandenburg2016} and SCAN functional \cite{Peng2015} for the xc energy, $E_{xc}$. See the S.I. for details.

\begin{figure}[bthp]
\centering
\includegraphics[width=1.\linewidth]{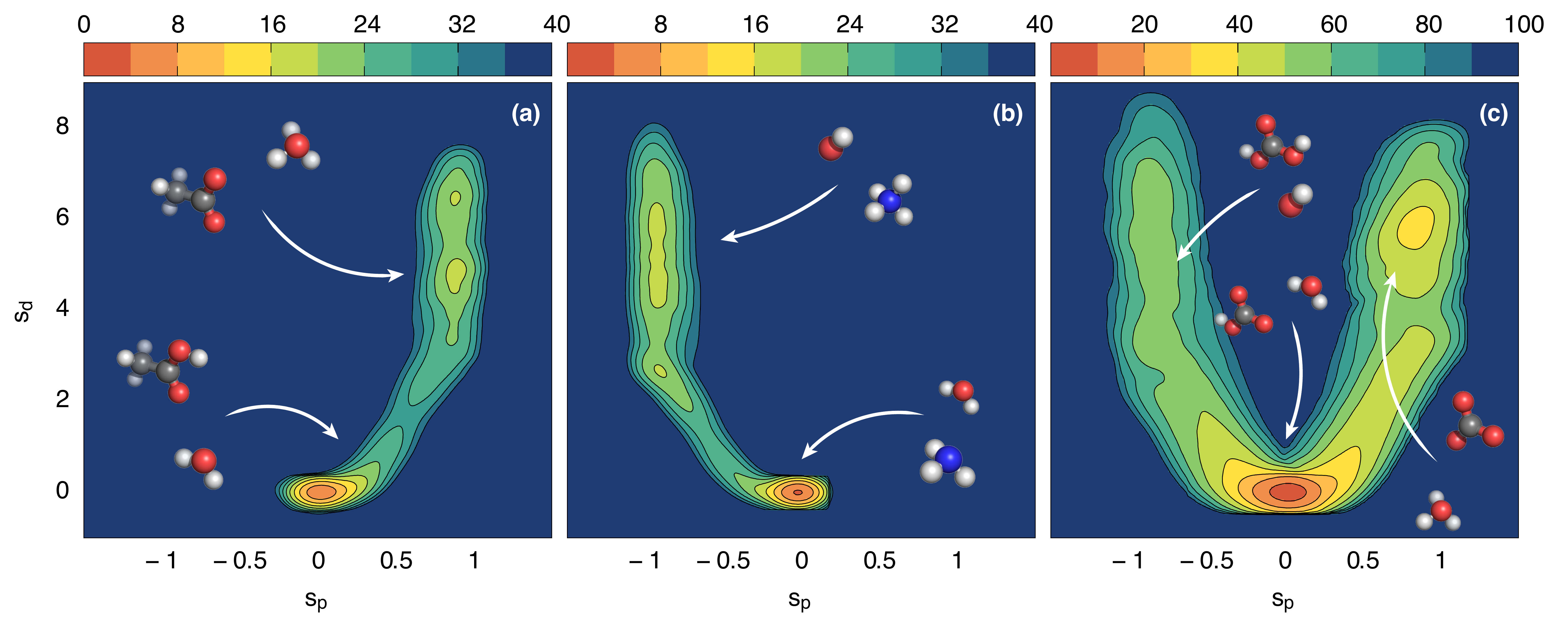}
  \caption{Free energy surfaces along $s_p$ and $s_d$ of acetic acid (a), ammonia (b) and bicarbonate (c) in aqueous solution. Colorbars indicate the free energy expressed in \si{kJ~mol^{-1}} units. The CV $s_d$ is expressed in \AA{}.}
  \label{fig:fes}
\end{figure}

In Fig.~\ref{fig:fes} we plot the Free Energy Surfaces (FESs) as a function of $s_p$ and $s_d$. These FESs vividly reproduce the expected behaviour. They all have a minimum at $s_p = 0$ that correspond to the state in which no charges are present in the solvent. In the acetic acid FES (Fig.~\ref{fig:fes}-a) a second minimum close to $s_p = -1$ reflects its acid behaviour. By contrast, the ammonia FES (Fig.~\ref{fig:fes}-b) shows a second minimum close to $s_p = 1$. 
The shape of ammonia and acetic acid FES are approximately related by a mirror symmetry reflecting their contrasting behaviour. Similarly the bicarbonate symmetric FES (Fig.~\ref{fig:fes}-c) mirror its amphoteric character.

As the conjugate pair is formed $s_d$ starts to assume positive values corresponding to the separation and diffusion of the conjugate pair. Compared to the undissociated state in which only $s_d=0$ is allowed, states where a conjugate pair is present show an elongated shape of the basins along this variable. This is caused by the diffusive behaviour of hydronium e hydroxide ion in solution that makes accessible a continuum range of distances. Moreover, along this CV we can observe a barrier around 1.5 corresponding to the breaking of the covalent bond between the hydrogen atom and the acid-base site. 

\section*{Conclusions}

The general applicability of this method to systems with different nature is an important step made in their understanding and description. The scheme can be extended to include quantum nuclear effects with the use of path integrals molecular dynamics\cite{Parrinello1984StudyKCl}. This would be of quantitative significance since for instance $pK_a$ values are affected by deuteration.
Moreover, the absence of assumptions or impositions about reactive candidates or reaction paths allows extending this method to systems of increasing complexity which cannot be addressed with traditional methods. Examples of questions that can now be answered are tautomeric equilibria in biochemical processes, acid behaviour in Zeolites and on the surface of oxides exposed to water.

\begin{acknowledgement}
This research was supported by the European Union Grant No. ERC-2014-AdG-670227/VARMET. Calculations were carried out on the ETH Euler cluster and on the M\"{o}nch cluster at the Swiss National Supercomputing Center (CSCS).
\end{acknowledgement}

%
\section{Supporting Information}
\subsection*{CV1: $\mathbf{s_p}$}
As described in the main text, the CV adopted to described the protonation state is
\begin{equation}
    s_p=\sum_{k=0}^{N^s-1}2^k\cdot q_k.
\end{equation}

Here, every component $q_k$ can assume a value equal to $-1$ for the $k$-th group whose chemical behaviour is acid, $+1$ for a basic one, and equal to zero for unreacted groups. Then, these values are summed with a different weight given by the power of two of the group index $k$. The prefactor $2^k$ allows to linearly combine a multiplet of values with a mathematical trick reducing the vector $\Vec{q}\equiv$($q_0$,$q_1$,\dots,$q_{N^s-1}$) in a single unambiguous scalar number.
Assuming we don't know anything about the reactivity of a system composed by 3 different groups able to react, a priori we cannot exclude any of its 7 different protonation states (see Tab.~\ref{tab:s1}).

\begin{table}[h]
    \centering
    \caption{The three components of the vector $\Vec{q}$ and the respective CV values.}
    $\begin{array}{rrr}
       q_0 & q_1 & q_2 \\
       \hline
        0 & 0 & 0  \\
        1 & -1 & 0  \\
        -1 & 1 & 0 \\
        0 & 1 & -1  \\
        0 & -1 & 1 \\
        1 & 0 & -1  \\
        -1 & 0 & 1  
    \end{array}
    \Longrightarrow
    \begin{array}{r}
        s_p \\
       \hline
        0 \\
         -1 \\
         1 \\
         -2 \\
         2 \\
         -3 \\
         3 
    \end{array}$
    \label{tab:s1}
\end{table}
As shown, all of these protonation states occupy different positions in the CV space without overlap among the states. This ensures the possibility to explore all of them starting from the most energetically accessible until the highest one in energy. Moreover, this approach allows to address systems in which multiple and unknown competitive reactions are present without beforehand fix the reactive pairs.

\subsection*{CV2: $\mathbf{s_d}$}

This CV returns a value proportional to the distance between the fully formed conjugate acid-base pair. Once the proton transfer has taken place, two acid-base sites will have an anomalous number of hydrogen atoms within their Voronoi polyhedra.
We can define the partial charge $\delta _i$ assigned to the $i$-th Voronoi polyhedron as

\begin{equation}
    \delta_{i} = W_{i \in k'} - \frac{N_{H \in k'}}{N_{k'}}, 
\end{equation}
where $N_{k'}$ and $N_H$ are constants indicating respectively the total number acid-base sites belonging to the $k'$-th group and the hydrogen atoms bonded to them in the equilibrium state, while $W_i$ is the instantaneous number of hydrogen atoms assigned to the $i$-th acid-base site.

\begin{figure}[htb]
    \centering
    \includegraphics[width=.8\textwidth]{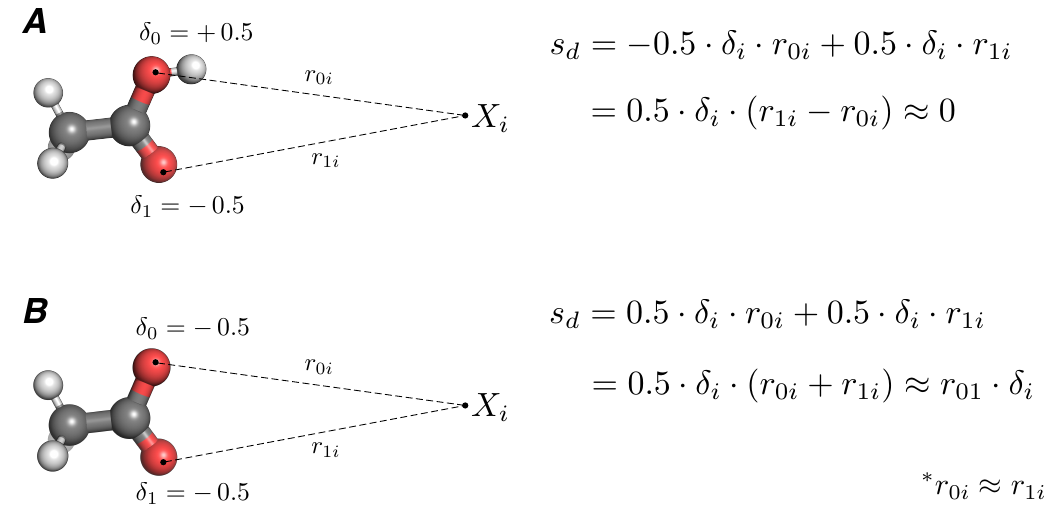}
    \caption{Schematic view of $s_d$ calculation between Acetic acid (A) or acetate (B), and a generic species $X_i$.}
    \label{fig:my_label}
\end{figure}

This means, for example, that in an system composed by a molecule of acetic acid and 31 molecules of water, every water oxygen atoms has $N_H=62$ and $N_0 = 31$ while the acid ones have $N_H=1$ and $N_1 = 2$. In the initial frame all the water sites have a values of $W_i$ close to 2, $N_H/N_0 = 2$ and therefore $\delta_i$ close to zero. After having subtracted a proton by the acid molecule, one of the sites $i$ of the solvent will assume a value of $W_{i'}$ close to 3 and $\delta_{i'}$ to 1. The two acetic acid oxygen atoms have only one hydrogen assigned to them and then $N_H/N_0 = 0.5$. In the undissociated species $W_i$ is 1 for the site bonded to the hydrogen atom and 0 to the other one making the $\delta_i$ values equal to +0.5 and -0.5 respectively. After the dissociation these sites must be indistinguishable and the acid molecule able to capture again a proton with one of them without any preference. Then, the partial charges will be -0.5 for both of the sites. This ensures that the opposite sign terms cancels each other in the undissociated case (Fig.~\ref{fig:my_label}-A) and gives an averaged contribution in the dissociated one (Fig.~\ref{fig:my_label}-B).

\subsection*{Restraint: $\mathbf{s_r}$}

\begin{figure}[htb]
    \centering
    \includegraphics[width=.5\textwidth]{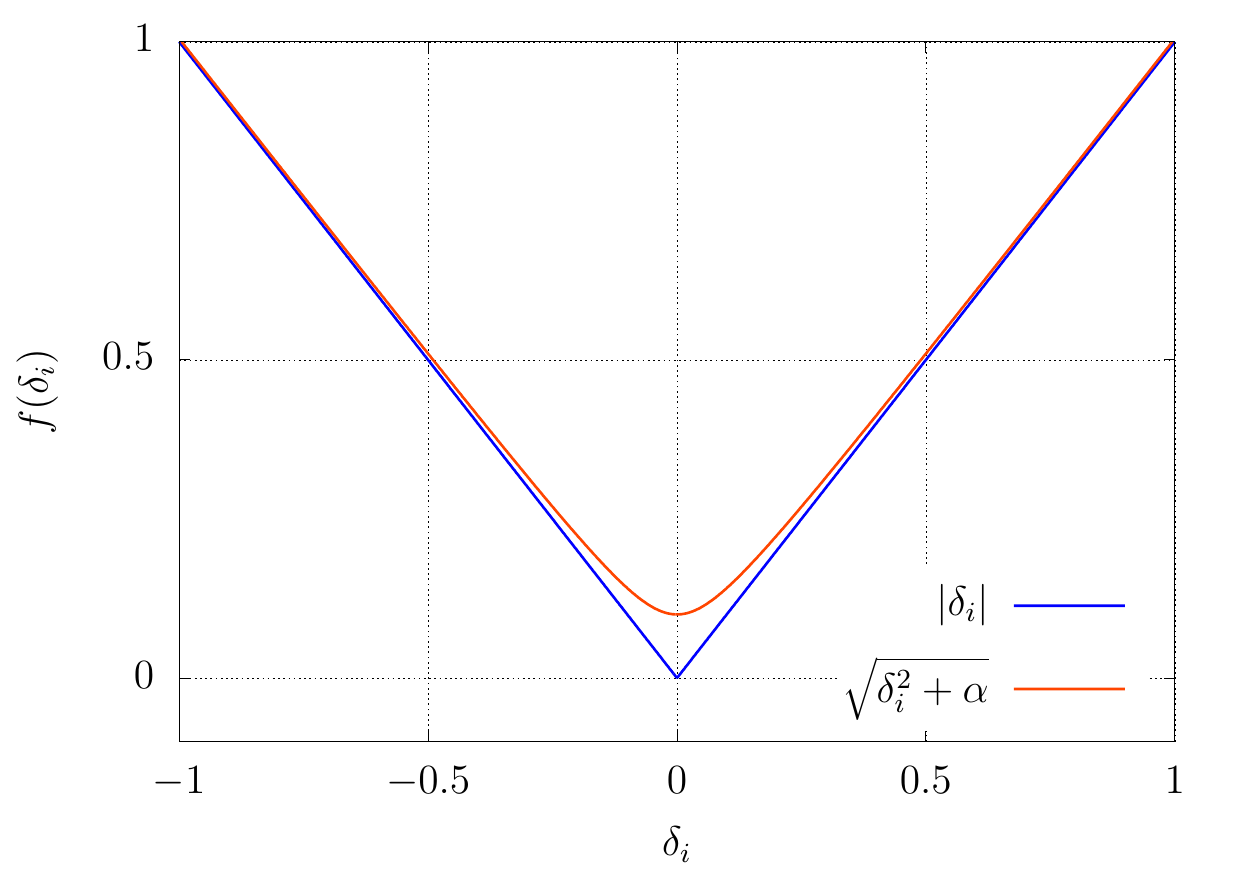}
    \caption{Different behaviour of absolute value function (blue line) and the smoothed version (orange line) in proximity of $\delta_i = 0$. The parameter $\alpha$ controls the smoothness of the curve. In this plot the value of $\alpha$ has been set equal to $10^{-2}$.}
    \label{fig:t}
\end{figure}

A restraint has been applied in order to avoid the formation of more then one conjugate acid-base pair.

\begin{equation}
    s_r = \sum_i \sqrt{\delta_i^2 + \alpha},
\end{equation}
where $i$ run all over the acid-base site indexes and $\alpha$ is a positive number much less than 1.
With a proper value of $\alpha$ the square root term is a good approximation of the absolute value that allows to avoid the singularity for $\delta_i = 0$ (see Fig.~\ref{fig:t})

This CV returns the summation of the all the partial charge moduli. This function can be restrained limiting at the given time the number of reacted pairs simultaneously present.

\subsection*{Ab initio MD setup}
As reported in the main text, all the simulations have been performed with Born-Oppenheimer MD simulations using CP2K package \cite{Vandevondele2005} patched with PLUMED2 \cite{Brandenburg2016}. Details and parameters adopted in the ab initio MD simulations are reported in Tab.~\ref{tab:my_label} . 

\begin{table}[h]
    \centering
    \caption{Ab initio MD parameters.}
    \begin{tabular}{rrrr}
    & \textbf{ACETIC ACID} & AMMONIA & BICARBONATE \\
       \hline
    Reactive molecule & 1 & 1 & 1 \\
    Water molecules & 31 & 31 & 31 \\
    Ensemble & NVT & NVT & NVT \\
    Temperaure (K) & 300 & 300 & 300 \\
    Thermostat & CSVR\cite{Bussi2007} & CSVR\cite{Bussi2007} &  CSVR\cite{Bussi2007} \\
    Cell length (\AA) & 9.97 & 9.80  & 10.47 \\ 
    Basis sets & TZV2P-GTH & TZV2P-GTH & TZV2P-GTH \\
    Potential & GTH-PBE & GTH-PBE & GTH-PBE \\
    Energy cutoff (Ry) & 600 & 600 & 800 \\
    Relative cutoff (Ry) & 60 & 60 & 80 \\
    EPS SCF & 1.0E-6 & 1.0E-6 & 1.0E-6 \\
    XC Functional & SCAN \cite{Peng2015} & SCAN \cite{Peng2015} & SCAN \cite{Peng2015} \\
    Time step (fs) & 0.5 & 0.5 & 0.5 \\
    Length time (ps) & 258 & 285 & 461 \\
    \end{tabular}
    \label{tab:my_label}
\end{table}

\subsection*{Box thermalization}

Each system, composed by 31 molecules of water and 1 of solute, has been thermalized as reported in Tab.~\ref{tab:my_label3}.

\begin{table}[htb]
    \centering
    \caption{Thermalization protocol.}
    \begin{tabular}{lllll}
    Step & Type & Ensemble & Length time (ps) & XC funct. \\
    \hline
    1 & Geom. Opt. & -   & -   & PBE\cite{Perdew1996}  \\
    2 & MD         & NVT & 1   & PBE\cite{Perdew1996}  \\
    3 & MD         & NPT & 10  & PBE\cite{Perdew1996}  \\
    4 & MD         & NVT & 2.5 & PBE\cite{Perdew1996}  \\
    5 & MD         & NVT & 2.5 & SCAN \cite{Peng2015} \\
    \end{tabular}
    \label{tab:my_label3}
\end{table}
The reason for this somewhat odd looking schedule is that the NPT ensemble module of CP2K does not support SCAN.

\subsection*{Well-tempered metadynamics setup}
Parameters adopted for PLUMED2 settings are reported in Tab.\ref{tab:my_label2}
\begin{table}[h]
    \centering
    \caption{PLUMED parameters.}
    \begin{tabular}{rrrr}
    & ACETIC ACID & AMMONIA & BICARBONATE \\
       \hline
    Gaussian hills heights & 0.25 & 0.25 & 0.5 \\
    Gaussian hills widths ($s_p$) & 0.2 & 0.2 & 0.2 \\ 
    Gaussian hills widths ($s_d$) & 0.4 & 0.4 & 0.4 \\ 
    Bias factor & 10 & 10 & 15 \\
    Temperature (K)& 300 & 300 & 300 \\
    Hills deposition rate & 100 & 100 & 100 \\
    $\lambda$ ($s_p$) & 5 & 5 & 5 \\
    $\lambda$ ($s_d$) & 8 & 8 & 8 \\
    $\lambda$ ($s_r$) & 12 & 12 & 12\\
    $\alpha$ ($s_r$) & 1.0E-4 & 1.0E-4 & 1.0E-4 \\
    \end{tabular}
    \label{tab:my_label2}
\end{table}


\bibliography{references}

\end{document}